\documentclass[12pt]{article}
\usepackage{epsfig}
\def\be{\begin{equation}}
\def\ee{\end{equation}}
\newcommand{\bea}{\begin{eqnarray}}
\newcommand{\eea}{\end{eqnarray}}

\newcommand{\bp}{{\bf p}}

\def\lsim{\mathrel{\rlap{\lower4pt\hbox{\hskip1pt$\sim$}}
    \raise1pt\hbox{$<$}}}         

\def\gsim{\mathrel{\rlap{\lower4pt\hbox{\hskip1pt$\sim$}}
    \raise1pt\hbox{$>$}}}         

\title{Corrections to the generalized vector dominance
due to diffractive $\rho_3$ production}

\date{}
\author{I.P. Ivanov$^{1,2}$\thanks{E-Mail: Igor.Ivanov@ulg.ac.be}, S. Pacetti$^{3}$, \\
  {\normalsize $^1$  Fundamental Interactions in Physics and Astrophysics Group, Universit\'{e} de Li\`{e}ge, Belgium} \\
  {\normalsize $^2$ Sobolev Institute of Mathematics,  Novosibirsk, Russia}\\
  {\normalsize $^3$ LNF INFN, Frascati, Italy}}

\begin{document}
\maketitle

\begin{abstract}
The idea of the vector dominance is still in use in various analyses of
experimental data of photon-hadron reactions.
It makes sense, therefore, to recast results of microscopic calculations
of such reactions in this language.
Here we present the diffractive DIS $\rho_3$ production as a specific correction
to the generalized vector dominance. We perform a coupled channel analysis of
spin-orbital excitations in diffractive photoproduction and reiterate the point
that $\rho_3$ in diffractive DIS will be sensitive to a novel aspect of diffraction.
\end{abstract}

\section{Introduction}

The study of photon-hadron collisions in 1960's was driven to large extent
by the Vector Dominance Model (VDM), the idea that
the photon in such reactions behaves as a universal combination of hadrons with the photon's
quantum numbers, see review \cite{Bauer}. In its simplest form, one assumes that
the ``hadronic part" of a physical photon in a given isospin-flavor channel
is saturated by the ground state vector meson $V$ contribution.
If accompanied with the assumption that the subsequent interaction of this meson
is a one-channel process, it yields direct relations among the cross
sections of different processes, such as $\sigma(\gamma p \to V p)$, $\sigma_{tot}(V p)$,
and $\sigma_{tot}(\gamma p)$ as well as decay width $\Gamma(V \to e^+e^-)$.
Lifting some of these restrictions has lead to Generalized Vector Dominance (GVD) models,
which provided rather good overall description of the data
on the medium energy photon-hadron interactions.

The advent of partonic description of high-energy reactions as well as
a vast amount of new data has set boundaries of the applicability of VDM/GVD.
A particularly transparent insight into the nature of vector dominance is
offered by the color dipole approach \cite{colordipole} (see next Section).
Still, the physically appealing idea behind the vector dominance
makes it an interesting exercise to recast results of a microscopic theory
in a VDM-like form. An example of such analysis was given in Ref.~\cite{NNNmultichannel}
where the photoproduction of the radially excited meson $\rho(2S)$
off nuclei was found to be due to the off-diagonal transitions among
different radial excitations in diffraction.
In a more recent example, \cite{E687-6pianalysis,crypto2007}, the GVD was used
to study the nature of a narrow dip structure in the $6\pi$ final state
located near $M_{6\pi}\sim 1.9$ GeV.

In this paper we discuss the recent results on diffractive $\rho_3$ production \cite{spin3},
obtained within the $k_t$-factorization approach, in the GVD language.
The $\rho_3(1690)$ meson cannot couple directly to the photon and therefore it is absent in the
annihilation $e^+e^- \to \gamma^* \to $ hadrons. But it can be produced
diffractively, since diffraction conserves only the $P$- and $C$-parities
but not the projectile spin $J$. Thus, $\rho_3$ production can be interpreted as
a specific correction to the vector dominance model.
With the coupled channel analysis we show that diffractive production
of the $D$-wave spin-1 and spin-3 mesons of the $\rho$ system,
despite having comparable cross sections, probe very different aspects
of diffraction.

The paper is organized as follows.
In Section 2 we discuss relation between the (generalized) vector dominance
models and the partonic description of diffraction.
In Section 3 we argue that the diffraction operator does not conserve
the spin of the projectile nor the angular momentum of the $q\bar q$ state,
which represents the projectile in the first approximation. Production of $\rho_3$, thus, can be viewed
as a result of the off-diagonal transitions between different hadronic states
in diffraction. In Section 4 we note that such a correction to VDM might have
already been observed by the E687 experimentally.
Possible nuclear effects and additional ``photophobic'' states
are discussed in Section 5. Finally, in Section 6 we draw our conclusions.

\section{(Generalized) Vector Dominance and its limits}

Let us first remind the standard assumptions behind VDM and discuss
the presence of excited mesons in the photon in this context.

In the original formulation, the physical photon
is represented as a sum of a bare photon and of a "hadronic" part of the photon.
Such decomposition is not Lorentz-invariant by itself, because
what appears as a hadronic part of the photon in one frame of reference
turns into a hadronic fluctuation of a target in another.
One usually chooses the target rest frame, and if the photon energy
is large enough, this decomposition is well defined.
It is the hadronic part of the photon that participate in hadronic
processes, while the bare photon contributes negligibly.

The hadronic part of the physical photon is represented as an integral
over all possible {\em asymptotic} (in respect to strong interactions)
hadronic states with photon's quantum numbers
and with invariant mass $M$.
At not too large masses, the dispersion integral over $M$ is saturated
by the lowest resonances. Such contributions can then be {\em defined}
as contributions of vector mesons. Limiting ourselves to the
flavor-isospin sector that corresponds to the $\rho$ mesons, one can
rewrite the hadronic part of the (virtual) photon as
\be
|\gamma^*(Q^2)\rangle_h = \sum_{V} {e\over f_V} {m_V^2 \over m_V^2 + Q^2}
| V\rangle\,.\label{fock1}
\ee
The simplest VDM consists in assumption that only the ground state meson dominates
in (\ref{fock1}), which leads to
$$
|\gamma^*\rangle_h = {e\over f_\rho} {m_\rho^2 \over m_\rho^2 + Q^2}
|\rho\rangle\,.
$$
This assumption is often accompanied with
an additional requirement that subsequent scattering
process is {\em diagonal} in the space of states $|V\rangle$ in (\ref{fock1}),
and it then leads to direct relations among various cross sections.

The presence of excited vector mesons in diffractive photoproduction
calls to lifting the above restrictions. In the Generalized Vector Dominance (GVD) model
one accepts (\ref{fock1}) as it is, and assumes further that the subsequent interaction
can lead to off-diagonal transitions among vector mesons $V_i\to V_f$.

\subsection{GVD in the color dipole language}

The origin of VDM/GVD success becomes transparent in the color dipole approach.
It applies to the frame where projectile momentum is large,
so that the transverse motion of partons is slowed down relativistically,
and the fact that individual partons are not asymptotic states becomes inessential.
In a high-energy diffractive reaction, the scattering amplitude has form
${\cal A}(A \to B) = \langle B|\hat \sigma|A\rangle$,
where diffractive states are represented as coherent combinations of multipartonic
Fock states:
\be
|A\rangle = \Psi^{A}_{q\bar q}|q\bar q\rangle +
\Psi^{A}_{q\bar q g}|q\bar q g\rangle + \dots\label{A-fock}
\ee
Here integration over all internal degrees of freedom assumed,
and $\hat\sigma$ is the diffraction operator than describes
the {\em diagonal} scattering of these multiparton states in the impact
parameter representation.
Switching from the basis of multipartonic states
to the basis of physical mesons $\{|V_i\rangle\}$ and assuming completeness,
one can recover (\ref{fock1}).

Due to the lowest Fock state domination, the diffraction operator
is based on the color dipole cross section $\sigma_{dip}(\vec r)$
of a $q\bar q$ pair with transverse separation $\vec r$.
The transition amplitude is represented as
\be
{\cal A}(A \to B) = \int dz d^2\vec r\ \Psi_{q\bar q}^{B *}(z,\vec r) \sigma_{dip}(\vec r)
\Psi_{q\bar q}^{A}(z, \vec r)\,,\label{colordipole}
\ee
where $z$ is the quark's fraction of the lightcone momentum of particle $A$.

The origin of the VDM success in reactions where $A$ is the hadronic part of the photon
lies in the fact that the typical wave functions of the ground state vector meson
used in phenomenology are very similar to the transverse photon lightcone wave function
at small $Q^2$.

As virtuality $Q^2$ grows, the $q\bar q$ wave function of the photon shrinks,
while the color dipole cross section behaves as $\sigma_{dip} \propto r^2$
at small $r$ and reaches a plateau at large $r$.
As a result, the function under integral (\ref{colordipole}),
where $A\equiv \gamma^*$ and $B$ is a ground state vector meson,
peaks at the scanning radius $r_S \sim 6/\sqrt{Q^2 + M^2}$, see Ref.~\cite{scanning}.
At small $Q^2$ the typical scanning radius is large,
and the amplitude is roughly proportional to the integration measure
$$
{\cal A}(\gamma^* \to V) \propto r_S^2 \propto {1 \over (Q^2 + M^2)}\,,
$$
which mimics the VDM behavior. At larger $Q^2$ the scanning radius becomes small enough
and the diffraction cross section itself {\em decreases}. This phenomenon
of color transparency produces a more rapid decrease
$A(\gamma^*\to V) \propto 1/(Q^2+M^2)^2$ up to logarithmic factors, \cite{colordipole,scanning}.

\subsection{Presence of excited vector mesons in the photon}\label{sectexcited}

The behavior just described can be cast in the GVD language involving radial excitations \cite{NNNmultichannel}.
At large $Q^2$, the (small) photon must be represented as a coherent combination
of a large number of (big) radially excited states.
Representing the diffractive production amplitude of a final meson $V$ as
$$
A(\gamma^*(Q^2)\to V) = \sum c_i(Q^2) {M_i^2 \over Q^2 + M_i^2} A(V_i \to V)\,,
$$
one sees that each term in this expansion decreases with $Q^2$ growth as $\propto 1/(Q^2 + M_i^2)$.
However, coefficients $c_i(Q^2)$ must behave in such a way
that cancellations among the terms makes the overall $Q^2$-dependence
of $A(\gamma^*\to V)$ is $\propto 1/(Q^2 + M_i^2)^2$, in accordance
with color dipole result.

Note that similar arguments must be at work for the large-$M$ photoproduction,
when one studies the large-mass tail of broad resonances in a given (for example, multipion) final state.
Production of a multipion state with invariant mass $M_{n\pi}$ significantly larger
than the nominal mass of the vector meson must involve $q\bar q$ pairs with larger invariant mass,
and smaller transverse separation, than for the vector meson at peak.
In the color dipole approach this effect can be roughly accounted for by an additional
correction factor
\be
F(M_{n\pi}) = {\sigma_{dip}(r_S(M_{n\pi})) \over \sigma_{dip}(r_S(M_{V}))} \label{correction}
\ee
in the amplitude. In the VDM language the same correction must be implemented as an additional
$M_{n\pi}$-dependence of the $\sigma(Vp\to Vp)$.

Another correction to VDM is related to the spinorial structure of the
hadron's coupling to the $q\bar q$ state, implicitly present in (\ref{A-fock})
in the definition of $\Psi_{q\bar q}^A$.
According to QED, the photon couples to the $q\bar q$ pair
as $\bar u \gamma^\mu u$, but the corresponding coupling of a vector meson
depends on $q\bar q$ angular momentum inside the meson.
For the pure $S$-wave and pure $D$-wave vector mesons the
structures $\bar u \Gamma^\mu u$ are \cite{in99}:
\be
\Gamma^\mu_S = \gamma^{\mu} + {2p^\mu \over M+2m}\,,\quad
\Gamma^\mu_D = \gamma^{\mu} - {4(M+m)p^\mu\over M^2-4m^2} \,.\label{spinorial}
\ee
Thus, the photon coupling represents a specific form of $S$-wave/$D$-wave
mixing:
\be
\gamma^\mu \Psi(q\bar q) = \Gamma^\mu_S \Psi_S(q\bar q) + \Gamma^\mu_D \Psi_D(q\bar q)\,,
\label{gammamu}
\ee
with appropriately normalized $\Psi_S(q\bar q)$ and $\Psi_D(q\bar q)$.
Since the $D$-wave vector meson can be approximated by the $q\bar q$ pair in the $L=2$ state,
this proves that decomposition (\ref{fock1}) must include orbitally excited vector
mesons as well. The partial width $\Gamma(\rho''\to e^+e^-)$ is known very poorly,
\cite{pdg}, which gives us only very rough estimate $1/f_{\rho''} \sim 0.2 (1/f_\rho)$,
which gives a 20\% contribution of the $D$-wave meson in (\ref{fock1}).
This value, however, supports the argument that the origin of $D$-wave state here is the quarks'
Fermi motion.

There are two competing mechanisms for diffractive production
of the orbitally excited vector mesons. First, the $D$-wave component
of the photon in (\ref{gammamu}) can get ``actuated" via diagonal scattering off
the target. The other mechanism involves off-diagonal transition of the $S$-wave part of (\ref{gammamu})
into the $D$-wave vector meson under the action of diffraction operator.
The $k_t$-factorization analysis of \cite{Dwave} did not specify which mechanism
was the dominant. The coupled channel analysis presented in the following Section will
help find the answer.

The same off-diagonal transitions that break the $q\bar q$ angular momentum conservation
and produce a $D$-wave vector meson can also produce its spin-orbital partner,
the $D$-wave spin-3 meson. Analysis of \cite{spin3} in the case of $\rho_3(1690)$
showed that its production rate is expected to be only $2\div 3$ times smaller than
production rate of $\rho''(1700)$, which is believed to be predominantly $D$-wave
vector meson. The hadronic part of the photon does not include the spin-3 meson,
so it arises exclusively due to the off-diagonal properties
of the diffraction operator.

\section{Coupled channel analysis of the orbital excitations in diffraction}

To get the GVD-like interpretation of the $\rho''(1700)$
and $\rho_3(1690)$ production, we perform a coupled channel analysis
of the action of diffraction operator in Fock subspace
generated by three states in the $\rho$ system:
the ground state meson $\rho(770)$, which we identify with the pure $1S$
state, the excited vector meson $\rho''(1700)$, which we identify
with a purely orbital excitation with $L=2$, and the spin-3 meson $\rho_3(1690)$,
which is also assumed to be in the $L=2$ state.

For numerical calculation of the transitions among these states in diffraction,
we use the color dipole/$k_t$-factorization representation of the production amplitude.
The dipole cross section is expressed in terms of
the unintegrated gluon distribution, whose fits were taken from \cite{in2000}
and adapted for the off-forward kinematics according to the procedure
described in \cite{review}.
The wave functions for the mesons were parametrized in the same
way as in \cite{Dwave} and \cite{spin3}. Note that the radial wave functions
of the three mesons considered do not have nodes. Therefore transitions from these states
to {\em radially} excited states (i.e. transitions away from this subspace) are weak.

We limit ourselves to the photoproduction ($Q^2=0$) and present cross sections
$\sigma_{ji} \equiv \langle j |\hat \sigma| i \rangle$ of transitions of an initial state $i$
with a given transverse polarization into final state $j$ with various polarization states.
The analytical expressions for the helicity amplitudes were derived according to the
standard diagram evaluation, see details in Appendix.

We start with the forward scattering, $d\sigma_{ji}/dt|_{t=0}$.
In this case the strict $s$-channel helicity conservation (SCHC) takes place,
and we are interested in transitions among transversely polarized states
of $\rho$, $\rho_D$ and $\rho_3$.
The calculations yield the following
matrix $d\sigma_{ji}/dt$ at $t=0$
\be
{d\sigma_{ji} \over dt}|_{t=0} = \left(
\begin{array}{ccc}
250 & 1.5 & 0.3 \\
1.5 & 460 & 1.3 \\
0.3 & 1.3 & 270
\end{array}\right) \ \mbox{mb$\cdot$GeV$^{-2}$}\,.\label{matrix1}
\ee
The off-diagonal values are non-zero, but stay small,
which means that both total spin $J$ and the $q\bar q$ angular momentum $L$
are not strictly conserved during diffraction.
We underline that the accuracy of our numerical results is low;
our calculations are plagued both by the large uncertainties of the
wave function parametrizations and of the experimental values of
$\Gamma(\rho''(1700)\to e^+e^-)$, which was used to fix the free parameters
of the $\rho_D$ and $\rho_3$ wave functions.
The off-diagonal elements in (\ref{matrix1}) show only the order of magnitude
of the effect; the error by a factor of 2--3 can be present.
The accuracy for the diagonal elements is somewhat higher, roughly
within $\sim 50\%$.

To obtain the integrated cross sections, we calculate
$d\sigma_{ji}/dt$ at non-zero $t$ and integrate it
within the region $|t| < 1$ GeV$^2$.
On passing to the non-forward cross sections, we must include
the helicity amplitudes transition that violate SCHC. Such transitions
give marginal contributions to the $L$-conserving
diagonal transitions, but they are expected to be more important
in the off-diagonal cases. In particular, results of \cite{spin3}
suggest that the $\rho_3$ production at small $Q^2$
can be even {\em dominated} by the helicity violating transitions.

Strictly speaking, in the non-forward case the diffraction operator
acts in the $3+3+7=13$-dimensional space of all helicity
states of these three mesons. To simplify presentation,
we show below the sum of cross sections
of transitions from a given transversely polarized initial state
to a final state with {\em all} possible helicities, which will
make the transition matrix non-symmetric.
The result of numerical integration is:
\be
\sigma_{ji} = \left(
\begin{array}{ccc}
19 & 1 & 0.2 \\
1 & 27 & 0.3 \\
1.3 & 0.4 & 19
\end{array}\right)\ \mbox{mb}\,.\label{matrix2}
\ee
Calculation showed that the diagonal elements
are mostly due to helicity conserving transitions, while the off-diagonal elements receive
very large contributions from helicity violating transitions, in agreement with expectations.
Note very large difference between $\sigma(\rho_S\to\rho_3)$
and $\sigma(\rho_3\to\rho_S)$, which also confirms domination of
helicity violating transitions in $\rho_3$ production.

\subsection{Difference in diffractive $\rho_D$ and $\rho_3$ production}

In order to understand the differences between photoproduction
of $\rho_D$ and $\rho_3$, consider the initial photon as a vector in the
subspace we consider. According to the discussion in Sect.~\ref{sectexcited},
it can be represented roughly as $|\gamma\rangle \sim |\rho_S\rangle + 0.2 |\rho_D\rangle$.
One sees that direct ``materialization'' of the $D$-wave component
of the photon followed by its diagonal scattering
has much larger amplitude than the $L$-changing transition
from the $S$-wave component ($0.2\cdot 27$ vs. $1$).

On the other hand, the $\rho_3$ must appear in diffraction via the
off-diagonal $L$- and $J$-violating elements of the
diffraction operator (\ref{matrix2}).
Thus, in contrast to the $\rho_D$, the $\rho_3$ production
probes {\em a novel aspect of diffraction}.

\section{Comparison between the $4\pi$ BaBar ISR and E687 data}

In this section we discuss if the corrections
to the GVD due to $\rho_3$ might have been already
observed in experiment.

The dominant decay channel of $\rho_3$ is $4\pi$ with branching ratio $BR(\rho_3 \to
4\pi) = 73\%$. Thus one can look for its presence in diffractive
photoproduction by comparing the rescaled E687 data \cite{E687} with BaBar initial state
radiation (ISR) data \cite{babar4pi} in $2(\pi^+\pi^-)$ final state.

Using GVD accompanied with the assumption that the diffraction operator is diagonal,
one obtains the following relation between
the $4\pi$ spectra in the $e^+e^-$ annihilation and photoproduction:
\be
{1 \over M_{4\pi}^2} {d\sigma(\gamma p \to 4\pi p) \over d M_{4\pi}}
\propto \sigma(e^+e^-\to 4\pi)\,.\label{naive}
\ee
The presence of $\rho_3$ in diffraction should manifest itself
as a bump in the photoproduction spectrum around $M_{4\pi} \sim 1.7$ GeV.
If the above ideas of the dominance of SCHC violation in $\rho_3$
are correct, one will see a larger bump at higher values of $|t|$.

\begin{figure}[!htb]
   \centering
\includegraphics[width=8cm]{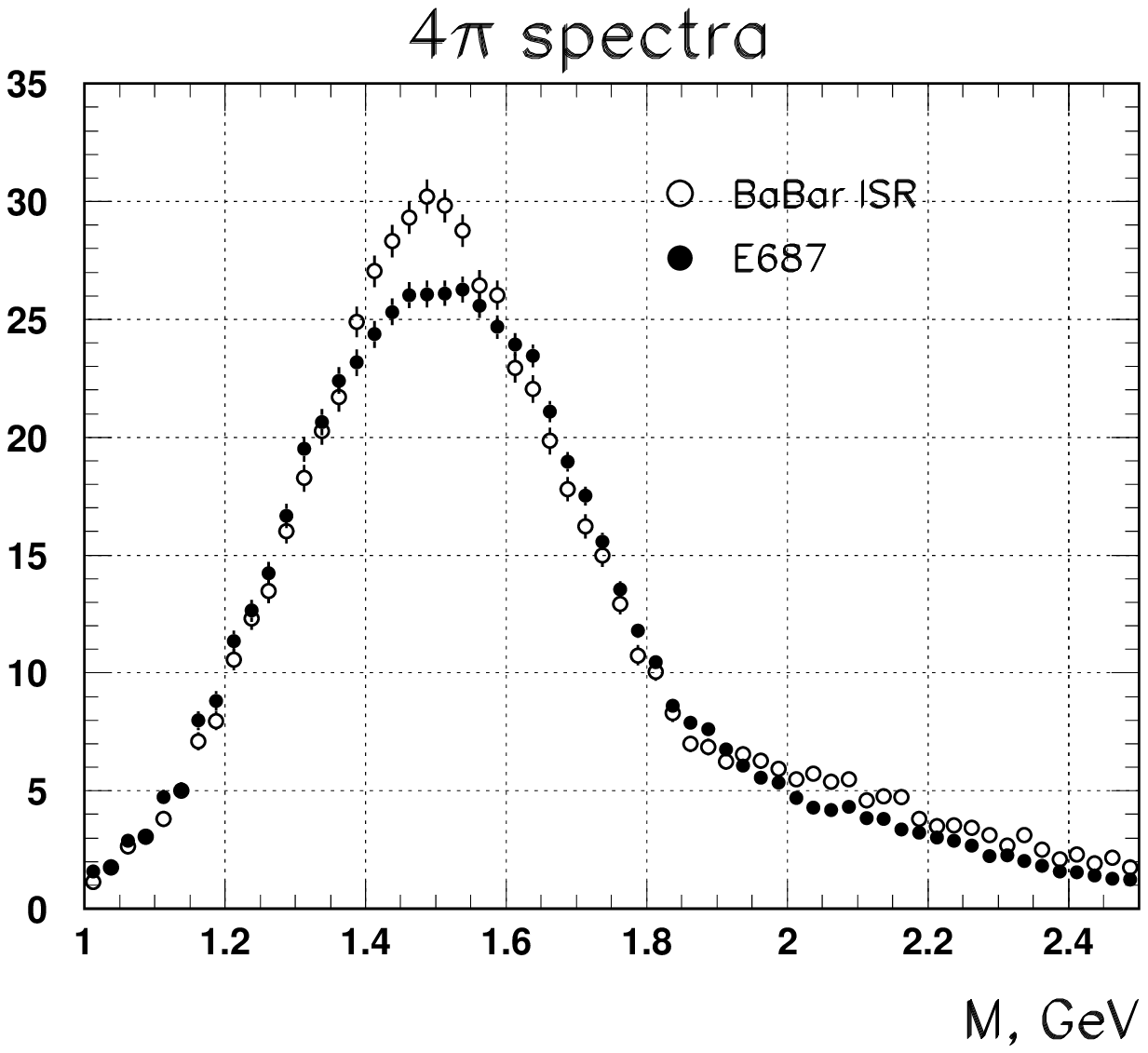}
\includegraphics[width=8cm]{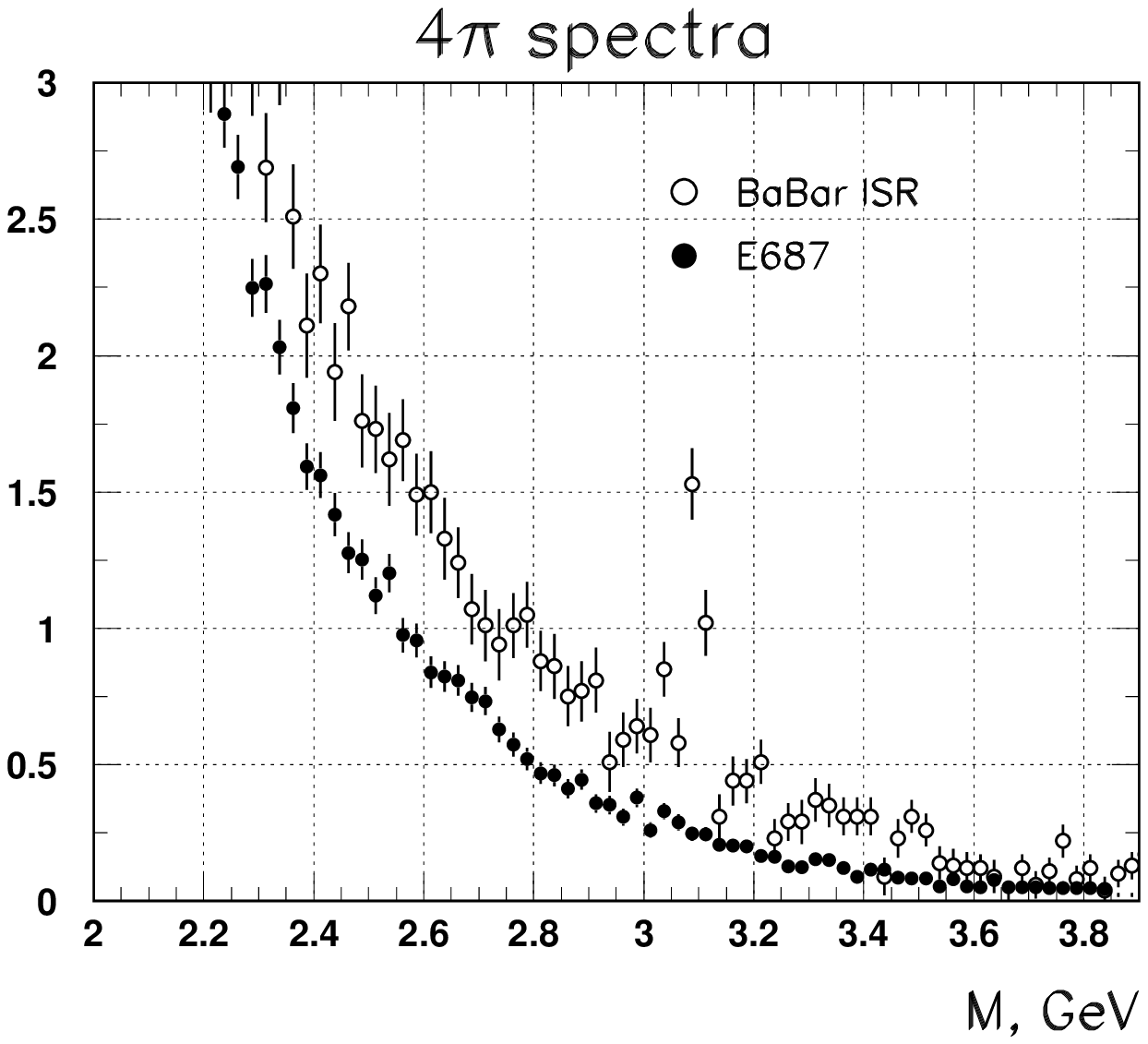}
\caption{Comparison between the BaBar data and the E687 data weighted
with $1/M^2_{4\pi}$ factor in the resonance region, left, and in the high mass region, right.}
   \label{fig-comp}
\end{figure}

In Fig.~\ref{fig-comp} we present the $4\pi$ spectrum in $e^+e^-$
annihilation obtained by BaBar and the diffractive
photoproduction cross section from E687 modified according to (\ref{naive}).
The relative normalization of the two data sets is adjusted manually
for a better comparison of the resonance peaks.

There are three regions where deviations are seen. At $M_{4\pi}\approx 1.5$ GeV
the BaBar data are significantly higher and at $M_{4\pi}\sim 1.7\div 1.8$ GeV
are somewhat lower than the rescaled E687 data. At $M_{4\pi}>2$ GeV
the BaBar data again take over. This region (zoomed in at the right plot of Fig.~\ref{fig-comp})
seems the most disturbing, not only because the ratio between the two data sets here
is large, but also because it increases with the $M_{4\pi}$ rise.

\begin{figure}[!htb]
   \centering
\includegraphics[width=8cm]{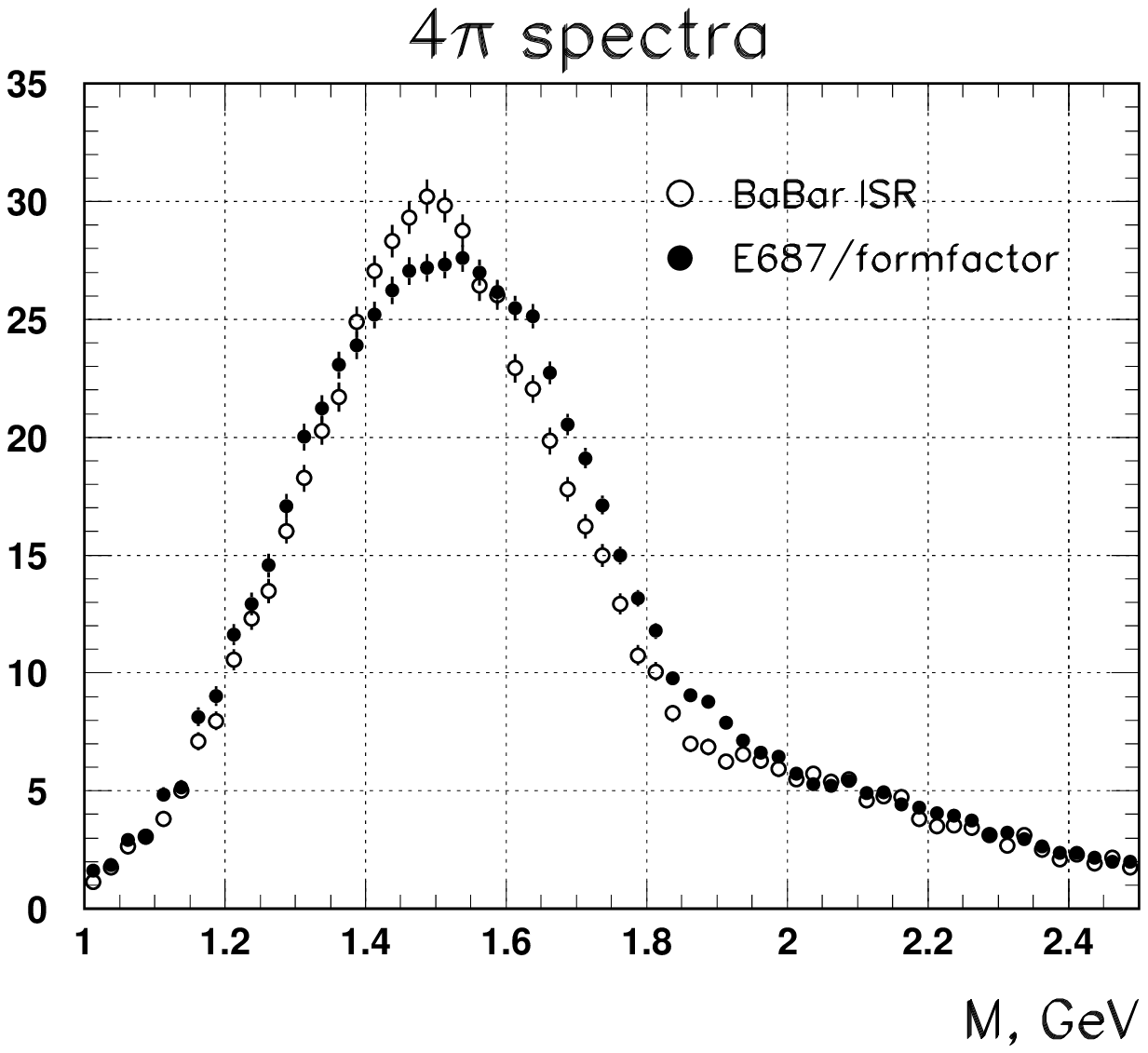}
\includegraphics[width=8cm]{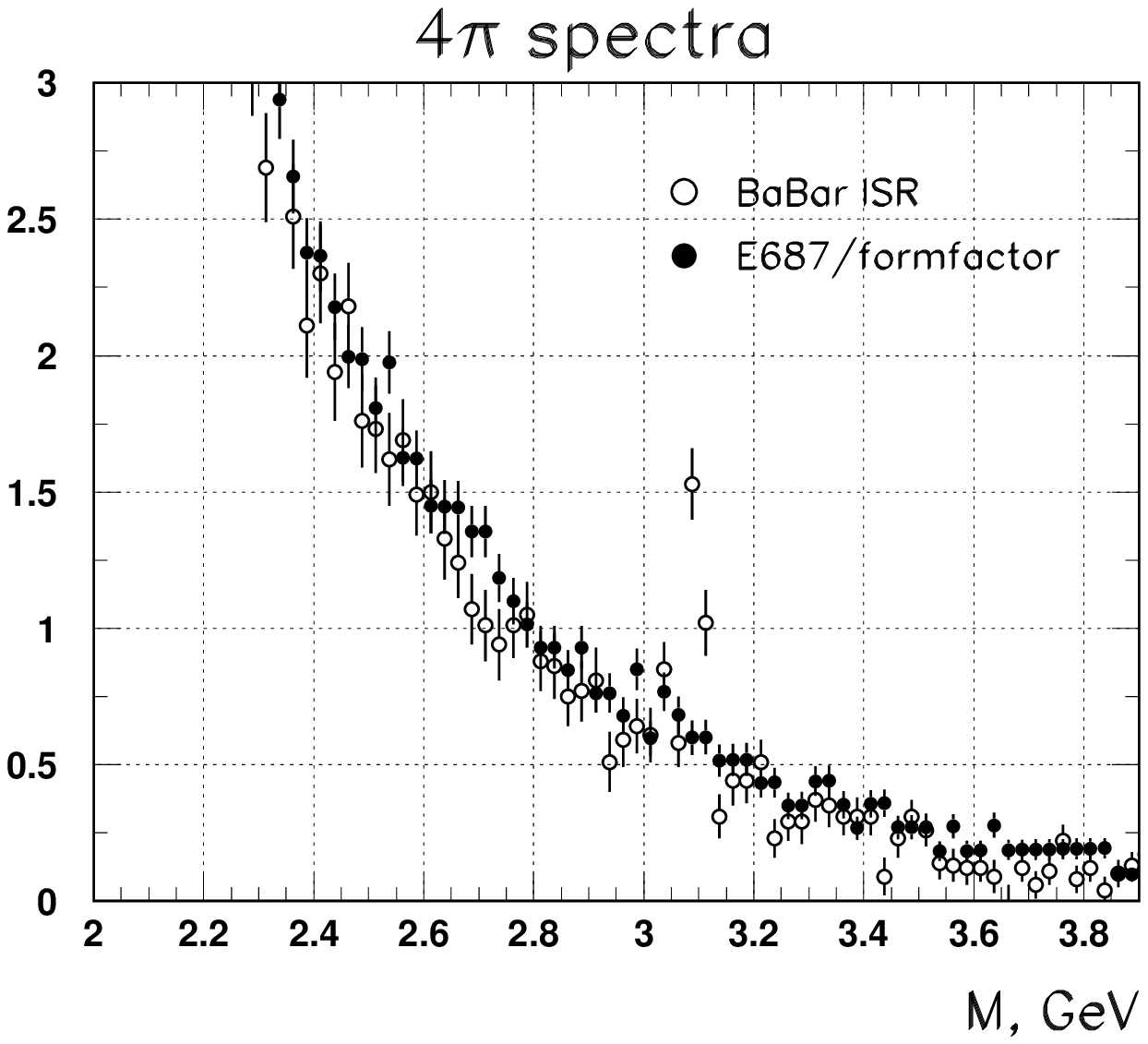}
\caption{The same as in Fig.~\ref{fig-comp} but with the E687 data additionally corrected
with the formfactor (\ref{ff}).}
   \label{fig-comp-2}
\end{figure}

We argue that this high-mass discrepancy is an artefact of
the naive VDM used in comparison (\ref{naive}). As discussed above,
diffractive production of high mass multipion states are additionally suppressed
in comparison with (\ref{naive}) by the factor (\ref{correction}).
In a phenomenological analysis,
this bias can be compensated by dividing the photoproduction data
by the square of the correction factor (\ref{correction}).
We used the well-known Golec-Biernat-W\"usthoff
saturation model \cite{GBW} for the color dipole cross section
$\sigma(r) = \sigma_0[1-\exp(-r^2/R^2(x))]$ and divided E687 data by
the additional compensation factor
\be
F(M_{4\pi}) = \left(1 - \exp\left[-{10\mbox{ GeV}^2 \over M_{4\pi}^2}\right]\right)^2\,,\label{ff}
\ee
and then readjusted the overall normalization.

Figure~\ref{fig-comp-2} shows the results. The simple factor (\ref{ff})
makes the two data sets nearly identical in the entire high-mass range shown,
$M_{4\pi} = 2.0$--3.9 GeV.
In the resonance region, the balance between the two experiments changes.
One sees a more prominent domination of the E687 data over
the BaBar data in the range of $M_{4\pi} \sim 1.6 \div 1.9$ GeV,
while the difference around $M_{4\pi} \approx 1.5$ GeV becomes less pronounced.

With these data sets only, one cannot draw a definitive conclusion about the origin
of the broad 1.6--1.9 GeV peak seen in the {\em difference} of the data sets.
It can be due to enhanced production of $\rho''(1700)$ or
due to the presence of $\rho_3$ in photoproduction.
If one assumes that its is {\em entirely} due to the presence of $\rho_3$,
one can roughly estimate its production rate,
\be
\sigma(\rho_3)/\sigma(\rho'+\rho'') \sim 0.05\div 0.1\,.\label{ratio}
\ee
This number appears to be in agreement both with
the old OMEGA result \cite{omega1986} and with calculations of \cite{spin3}.
We do not plunge here into a detail systematic analysis of the difference
of the two data sets, but just state that it is worth studying further.

The easiest way to resolve the ambiguity in the origin of the enhancement would be to measure
the same photoproduction spectrum at larger values of $|t|$ up to 1 GeV$^2$.
If $\rho_3$ photoproduction is indeed dominated by the helicity-flip amplitudes,
as argued in \cite{spin3}, its contribution should rapidly grow with $|t|$.

\section{Discussion}

\subsection{Nuclear effects}

A place where corrections to the naive VDM come to the foreground
is diffractive production of excited mesons off nuclei.
In this case the diffractive system can experience multiple
scattering off separate nucleons, which amounts to multiple action of the
diffraction operator on the initial state.
Such action enhances the rate of production of excited states
that were initially (almost) orthogonal to the photon.
The fingerprints of this effect in experiment would be
an observation of an $A$-dependence of the relative production rate of excited
states, the modifications of the shape of these resonances and, possibly,
novel interference patterns inside the nucleus.

Such in-medium modifications of the properties of
the radially excited $\rho$ states were explored in \cite{NNNmultichannel}.
Even at moderate energies the shape of the $\rho(2S)$ state
was noticeably distorted in heavy nuclei. The origin of this effect
was traced back to non-trivial interplay between two production mechanisms:
direct production $\gamma \to \rho(2S)$ and sequential transition
$\gamma \to \rho(1S) \to \rho(2S)$. The latter transition is precisely due
to the off-diagonal matrix element of the diffraction operator.

Similar effects are expected to take place in the orbitally excited sector
of the diffractive states.
In order to observe better the $\rho_3$, one must focus not at the forward production, but at the entire
region $|t| \lsim 1$ GeV$^2$.
As was discussed above, the $\rho_3$ production is exclusively due to the off-diagonal
matrix elements of the diffraction operator.
Besides, according to (\ref{matrix2}), transitions from $\rho_3$ back to the $\rho_S$
are less probable than the $\rho_S\to\rho_3$ transitions.
All this produces a persistent "flux" towards the $\rho_3$ state,
and its presence is enhanced upon each successive rescattering.

Note in addition, that production of $\rho_3$ in a given helicity state
can proceed via many different helicity sequences, such as
$\rho_S(\lambda_S) \to \rho_3(\lambda_3^\prime) \to \rho_3(\lambda_3)$.
All of them will interfere and might produce nontrivial patterns.

\subsection{Photophobic states in diffraction}

The $\rho_3$ is a state
whose direct coupling to the photon is zero (``photophobic'' state),
yet it appears among diffractive states due to the off-diagonal transition.
Similarly, one might expect that other hadrons
not coupled directly to the photon might show up in diffraction.
One interesting example is a hybrid meson.
Phenomenologically, one often treats the hybrid (vector) meson
as a state that does not couple directly to the photon,
but it can reappear in photon's Fock state decomposition
via hadronic loops and intermediate transitions to the nonexotic mesons.
An analysis of this type was performed in \cite{E687-6pianalysis,crypto2007}. There,
such a cryptoexotic state was assumed to couple to $\rho''(1700)$ but not to the photon.
This simple model was proposed to explain the narrow dip structure
in the $6\pi$ final state around $M_{6\pi}=1.9$ GeV observed both
in diffractive photoproduction \cite{E687-6pi} and in $e^+e^-$ annihilation \cite{babar4pi}.

The present coupled channel analysis seems to be a more adequate
framework for the analysis of possible interference effects
of such photophobic states in diffraction.
What one needs in order to get concrete predictions
is a (phenomenological) microscopic model for such a state.
Such an analysis would be
complementary to that of \cite{E687-6pianalysis,crypto2007},
since in these works the diffraction operator was assumed to be diagonal,
while we show that this assumption is unwarranted.
It would be interesting to see how non-diagonal transitions of the diffraction operator
influence the results of \cite{E687-6pianalysis,crypto2007}.

\section{Conclusions}

Since the vector dominance idea is still used these days to understand
some features of new experimental results, it is useful
to discuss the results of microscopic QCD calculations in the language of the
generalized vector dominance models. In this paper, we argued
that the vector dominance model, when applied to the region $M \sim 1.5-2.0$ GeV,
must receive significant corrections due to presence of the $\rho_3$
among diffractive states.

We compared the paths that lead to diffractive production of
$\rho''(1700)$, which is believed to be a $D$-wave vector meson,
and of $\rho_3(1690)$, its spin-orbital partner.
Recent $k_t$-factorization results \cite{spin3}
show that their cross sections should be comparable.
However, the coupled channel analysis performed here gives strong evidence that
these two processes probe very different aspects of the diffraction.
The $\rho''(1700)$ production can be viewed primarily as ``materialization''
of the $D$-wave component of the photon followed by diagonal diffractive scattering,
while the $\rho_3$ production probes exclusively the off-diagonal elements
of the diffraction operator.
Thus, with $\rho_3$ one can study novel aspects of diffraction.

We also compared recent E687 and ISR BaBar data on $4\pi$ spectra obtained
in diffraction and $e^+e^-$ annihilation, respectively, and
observed an enhancement in the photoproduction
precisely where $\rho_3$ resides. At present it is not known
if this enhancement is due to excited vector mesons or to the $\rho_3$, but
studies at non-zero momentum transfer $t$ might provide the answer.

Finally, we discussed the role of orbital excitations in photon-nuclear collisions,
and argued that the coupled channel analysis might help
study other ``photophobic'' states.
\\

{\bf Acknowledgements}. The work was supported by FNRS
and partly by grants RFBR 05-02-16211 and NSh-5362.2006.2.

\appendix
\section{Transition amplitudes}

A generic amplitude of diffractive transition of an initial meson with polarization $\lambda_i$
into the final meson with polarization $\lambda_f$ is written within the $k_t$-factorization approach
as
\be
{1\over s}Im\, A_{\lambda_f\lambda_i} =  {c_V \sqrt{4\pi\alpha_{em}} \over 4 \pi^2}
\int {dzd^2\vec k\over z^2(1-z)^2}  \int {d^2\vec \kappa \over \vec \kappa^4}
\alpha_s\,{\cal F}(x_1,x_2,\vec \kappa,\vec\Delta)\cdot
\sum_{\mathrm{diagr.}}I_{\lambda_f;\lambda_i}\Psi_f^*(\bp_2^2)\Psi_i(\bp_1^2)\,.\label{ampl}
\ee
Here $z$ is the lightcone momentum fraction of the quark, $\vec k$ is the relative
transverse momentum of the $q\bar q$ pair, while
$\vec \kappa$ is the transverse momentum of the gluon.
Coefficient $c_V$ is the flavor-dependent average charge
of the quark, the argument of the strong coupling constant
$\alpha_s$ is max$[z(1-z)(Q^2+M^2),\vec\kappa^2]$,
and ${\cal F}(x_{1},x_{2},\vec \kappa,\vec\Delta)$
is the skewed unintegrated gluon distribution,
with $x_1 \not = x_2$ being the fractions of the proton's momentum
carried by the uppermost gluons, see details in \cite{review}.
The wave functions of the initial and final meson depend on
\be
\bp_i^2 = {1\over 4}(M_i^2 - 4m_q^2) = {1\over 4}M_i^2(2z-1)^2 + \vec k_i^2 = k_{iz}^2 + \vec k_i^2\,,\quad i=1,2\,,
\label{bpi}
\ee
where $M_i$ is the invariant mass of the initial ($i=1$) and final ($i=2$) $q\bar q$ pair.
The integration variable $\vec k$ is taken equal to the final transverse momentum, $\vec k = \vec k_2$,
while the initial relative $q\bar q$ momentum $\vec k_1$ changes from one diagram to another.

The integrands $I_{\lambda_f\lambda_i}$ depend on the mesons considered.
For the $S$-wave to $S$-wave transition the integrands have form:
\bea
I^{SS}_{00} & = & {1 \over 4}M_1 M_2 \left[A_1 A_2 + {4(\vec k_1\vec k_2)(2z-1)^2 \over (M_1+2m_q)(M_2+2m_q)}\right]\,,\label{SS}\\
I^{SS}_{++} & = & (\vec k_1\vec k_2) + m_q^2 \left[B_1 B_2 + {4(\vec k_1\vec k_2)(2z-1)^2 \over (M_1+2m_q)(M_2+2m_q)}\right]\,,\nonumber\\
I^{SS}_{0+} & = & {1\over 2}(2z-1)M_2 \left[k_{2+}{2m_q \over M_2+2m_q} -
k_{1+}{2m_q \over M_1+2m_q}A_2 + k_{1+}{4(\vec k_1\vec k_2)\over (M_1+2m_q)(M_2+2m_q)}\right]\,,\nonumber\\
I^{SS}_{+0} & = & {1\over 2}(2z-1)M_1 \left[k_{1+}^*{2m_q \over M_1+2m_q} -
k_{2+}^*{2m_q \over M_2+2m_q}A_1 + k_{2+}^*{4(\vec k_1\vec k_2)\over (M_1+2m_q)(M_2+2m_q)}\right]\,,\nonumber\\
I^{SS}_{-+} & = & k_{1+}k_{2+}\left[1 - {4m_q^2(2z-1)^2 \over (M_1+2m_q)(M_2+2m_q)}\right]
- (k_{1+})^2{2m_q \over M_1+2m_q}B_2 - (k_{2+})^2{2m_q \over M_2+2m_q}B_1\,,\nonumber
\eea
and the remaining integrands can be obtained by appropriate change of $+$ to $-$ together with factor $(-1)^{\lambda_i+\lambda_f}$.
Here $k_{i\pm} = - (k_{i\mu} e_\pm^{\mu}) = - k^*_{i\mp}$, and
$$
A_i = 4z(1-z) + {2m_q \over M_i+2m_q}(2z-1)^2\,,\quad
B_i = 1+{\vec k_i^2 \over m_q(M_i+2m_q)}\,.
$$
Corresponding expressions for all other possible transitions among $S$-wave, $D$-wave and spin-3 states
can be obtained by the projection technique described in \cite{in99,spin3}.
For example, the corresponding integrands for the spin-3 meson transition from polarization state $\lambda_i$ to $\lambda_f$
can be described by $7\times 7$ matrix:
\be
I^{33}_{\lambda_f\lambda_i} = T^{3S}_{\lambda_f\lambda'}I^{SS}_{\lambda'\lambda}T^{S3}_{\lambda\lambda_i}\,,
\ee
where ``transition matrices" can be readily constructed from the Clebsch-Gordan coefficients
involved in description of the spin-3 meson, see Ref.~\cite{spin3}. For example,
\be
T^{S3}_{\lambda\lambda_i} =
\left(\begin{array}{ccccccc}
k_+^2 & {2\over \sqrt{3}}k_zk_+ & {1 \over\sqrt{15}}(2k_z^2-\vec k^2) & {2\over\sqrt{10}}k_zk_- & {1\over \sqrt{15}}k_-^2 & 0 & 0\\
0 & {1\over \sqrt{3}}k_+^2 & {4 \over\sqrt{15}}k_z k_+ & {1\over\sqrt{10}}(2k_z^2-\vec k^2) & {4 \over\sqrt{15}}k_z k_- & {1\over \sqrt{3}}k_-^2 & 0 \\
0 & 0 & {1\over \sqrt{15}}k_+^2 & {2\over\sqrt{10}}k_zk_+ & {1 \over\sqrt{15}}(2k_z^2-\vec k^2) & {2\over \sqrt{3}}k_zk_- & k_-^2
\end{array}
\right)\,,
\ee
where subscript $1$ is assumed for all the momenta, while matrix $T^{3S}$ is just the hermitian conjugate of $T^{S3}$ with replacement
$k_1 \to k_2$. Similar expressions can be obtained also for the $D$-wave vector mesons.

\end{document}